\documentclass[twocolumn]{revtex4}
\usepackage{amssymb}
\usepackage{latexsym}
\usepackage{epsfig}
\usepackage{color}
\usepackage{float}
\usepackage{graphicx}
\usepackage{epstopdf}

\begin{document}

\title{Accelerated cosmos in a non-extensive setup}
\author{H. Moradpour$^1$\footnote{h.moradpour@riaam.ac.ir}, Alexander Bonilla$^2$\footnote{abonilla@fisica.ufjf.br}, Everton M. C. Abreu$^{2,3}$\footnote{evertonabreu@ufrrj.br} and Jorge Ananias Neto$^2$\footnote{jorge@fisica.ufjf.br}}
\address{$^1$ Research Institute for Astronomy and Astrophysics of Maragha (RIAAM), Maragha 55134-441, Iran\\
$^2$ Departamento de F\'isica, Universidade Federal de Juiz de Fora, 36036-330, Juiz de Fora, MG, Brazil\\
$^3$ Grupo de F\'isica Te\'orica e Matem\'atica F\'isica,
Departamento de F\'isica, Universidade Federal Rural do Rio de
Janeiro, 23890-971, Serop\'edica, RJ, Brazil}

\begin{abstract}
Here, we consider a flat FRW universe whose its horizon entropy
meets the R\'{e}nyi entropy of non-extensive systems. In our
model, the ordinary energy-momentum conservation law is not always
valid. By applying the Clausius relation as well as the Cai-Kim
temperature to the apparent horizon of a flat FRW universe, we
obtain modified Friedmann equations. Fitting the model to the
observational data on current accelerated universe, some values
for the model parameters are also addressed. Our study shows that
the current accelerating phase of universe expansion may be
described by a geometrical fluid, originated from the
non-extensive aspects of geometry, which models a varying dark
energy source interacting with matter field in the Rastall way.
Moreover, our results indicate that the probable non-extensive
features of spacetime may also be used to model a varying dark
energy source which does not interact with matter field, and is
compatible with the current accelerated phase of universe.
\end{abstract}

\maketitle

\section{Introduction}

The violation of energy-momentum conservation law in curved
spacetime has firstly been proposed by P. Rastall to modify
general relativity theory of Einstein (GR) \cite{rastall}. After
his pioneering work, various type of modified gravity in which
matter fields are non-minimally coupled to geometry have been
proposed \cite{cmc,cmc1,cmc2,genras,PRL,shabprl,shabprl1}. It has
been shown the Rastall correction term to the Einstein field
equations can not describe dark energy meaning that a dark
energy-like source is needed to model the current phase of
universe expansion in this framework \cite{prd}. But, if one
generalizes this theory in a suitable manner, then the mutual
non-minimal coupling between geometry and matter field may be
considered as the origin of the current accelerating phase and the
inflation era \cite{genras}. More studies on Rastall theory can be
found in Refs.
\cite{gc1,al1,al2,al3,smal,Rlag,obs1,rastbr,rascos1,rasch,more1,more2,more3,more4,more5,hm,msal,plb1c,msh,clm,net}.

In Einstein gravity, horizons may meet the Bekenstein-Hawking
entropy-area law which is a non-extensive entropy
\cite{rn1,rn2,rn4,rn5,rn6,rn7,rn8,rn9,rn10,rn11,rn12,rn13,rn14,rn15,rn16,rey2}.
Moreover, it has recently been argued that a deep connection
between dark energy and horizon entropy may exist in gravitational
theories
\cite{cana,cana1,mmg,em,mitra,mitra1,mitra2,mitra3,mms,md,msgj,mswr}.
Indeed, although extensive statistical mechanics and its
corresponding thermodynamics lead to interesting results about the
universe expansion history \cite{mrej}, the mentioned points
encourage physicists to use non-extensive statistical mechanics
\cite{reyo1,tsallis} in order to study the thermodynamic
properties of spacetime and its related subjects
\cite{rn3,rey1,rsal1,rsal2,rsal,anp,ijtp,cite5,Kom1,Kom2,Kom3,Kom,EPJC,MSK}.

Recently, applying R\'{e}nyi entropy to the horizon of FRW
universe and considering a varying dark energy source interacting
with matter field, N. Komatsu found out modified Friedmann
equations in agreement with observational data on the current
phase of universe expansion \cite{EPJC}. Therefore, in his model,
total energy-momentum tensor including the matter field and
varying dark energy-like source is conserved. Combining this
entropy with entropic force scenario, one can also obtain a
theoretical basis for the MOND theory \cite{MSK}. In fact, the
probable non-extensive features of spacetime may be considered as
an origin for both the MOND theory and the current accelerated
expansion phase in a universe filled by a pressureless source
satisfying ordinary conservation law \cite{MSK}. Finally, it is
useful to note here that both mentioned attempts \cite{EPJC,MSK}
used the Padmanabhan holographic approach \cite{Padd} in getting
their models of the universe expansion.

Here, we are interested in obtaining a model for the universe
dynamics by applying the Clausius relation as well as the
R\'{e}nyi entropy to the horizon of FRW universe which has
non-minimally been coupled to matter field. Therefore, total
energy-momentum tensor does not necessarily satisfy the ordinary
conservation law, and in fact, it follows the Rastall hypothesis
in our setup.

The paper is organized as follows. In the next section,
introducing our approach, we present a thermodynamic description
for Friedmann equations in Rastall theory. Using the R\'{e}nyi
entropy, our model of universe is obtained in
Sec.~($\textmd{III}$). In the fourth section, we consider a
universe filled by a pressureless source, and show that, in our
formalism, it can experience an accelerated expansion.
Sec.~($\textmd{V}$) includes the observational constraints of
model. The last section is devoted to a summary and concluding
remarks. We use the unit of $c=\hbar=k_B=1$ in our calculations.


\section{Thermodynamic description of Friedmann equations in Rastall theory}

Based on the Rastall hypothesis \cite{rastall}

\begin{eqnarray}\label{rastal}
T^{\mu}_{\ \nu\ ;\mu}=\lambda R_{,\nu},
\end{eqnarray}

\noindent where $\lambda$ denotes the Rastall constant parameter,
and $T^{\mu \nu}$ is the energy-momentum tensor of source which
fills background. Moreover, $R$ is the Ricci scalar of spacetime.
This equation says that there is an energy exchange between
spacetime and cosmic fluids due to the tendency of geometry to
couple with matter fields in a non-minimal way \cite{genras,clm}.
For example, the $\lambda=\frac{1}{4k}$ case, where $k$ is called
the Rastall gravitational coupling constant, can support the
primary inflationary era in an empty FRW universe \cite{genras}.
In this manner, the ability of geometry to couple with matter
fields in a non-minimal way generates a constant energy density
equal to $\lambda R$ \cite{genras}. Some features of this
non-minimal mutual interaction and its corresponding energy flux
as well as its applications in cosmic eras have been studied in
Ref.~\cite{genras}.

Bearing the Bianchi identity in mind and integrating
Eq.~(\ref{rastal}), one can reach at \cite{rastall}

\begin{eqnarray}\label{r1}
G_{\mu \nu}+k\lambda g_{\mu \nu}R=k T_{\mu \nu}.
\end{eqnarray}

The Newtonian limit of the Rastall theory implies \cite{msal,msh}

\begin{eqnarray}\label{k}
k=\frac{\gamma}{\lambda}=\frac{4\gamma-1}{6\gamma-1}8\pi G,
\end{eqnarray}

\noindent where $\gamma\equiv k\lambda$. Applying the
thermodynamic laws to the horizon of spacetime and using
Eqs.~(\ref{r1}) and~(\ref{k}), it is shown that the horizon
entropy is achieved as \cite{hm,msal,plb1c}

\begin{eqnarray}\label{entf}
S_A^R=\frac{2\pi A}{k},
\end{eqnarray}

\noindent in which $A$ is the horizon area. Combining this result
with Eq.~(\ref{k}), one can easily find that entropy is positive
whenever $\gamma$ either meets the $\gamma<\frac{1}{6}$ or
$\gamma>\frac{1}{4}$ condition \cite{msal}. This equation can also
be written as $S_A^R=\frac{6\gamma-1}{4\gamma-1}S_B$, where
$S_B=\frac{A}{4G}$ is the Bekenstein-Hawking entropy \cite{msal},
meaning that the second law of thermodynamics is obtained for the
mentioned values of $\gamma$, if it is satisfied by the
Bekenstein-Hawking entropy. In addition, simple calculation lead
to

\begin{eqnarray}\label{lam}
\lambda=\frac{\gamma}{k}=\frac{\gamma(6\gamma-1)}{(4\gamma-1)8\pi
G},
\end{eqnarray}

\noindent for the Rastall constant parameter \cite{msal}. Indeed,
Eqs.~(\ref{k}) and~(\ref{lam}) are the direct results of imposing
the Newtonian limit on Eq.~(\ref{r1}), indicating that only for
$\lambda=\gamma=0$ we have $k=8\pi G$ \cite{rastall,msh}. In
Ref.~\cite{net}, assuming $k=8\pi G$ and studying Neutron stars in
Rastall gravity, authors found out that $\lambda$ is so close to
zero. Therefore, we see that since they consider $k=8\pi G$, their
result does not reject the Rastall hypothesis. Finally, it is
worth to remind here that the Bekenstein-Hawking entropy
($S_B=\frac{A}{4G}$) is also obtainable at the appropriate limit
of $\gamma\rightarrow0$.

For a flat FRW universe with scale factor $a(t)$ and line element

\begin{eqnarray}\label{rw}
ds^2=-dt^2+a(t)^2[dr^2+r^2(d\theta^2+\sin(\theta)^2d\phi^2)],
\end{eqnarray}

\noindent the apparent horizon, equal to the Hubble horizon, is
located at

\begin{eqnarray}\label{ah}
\tilde{r}_A=a(t)r_A=\frac{1}{H},
\end{eqnarray}

\noindent and therefore $A=\frac{4\pi}{H^2}$. Now, if the geometry
is filled by a prefect fluid with energy density $\rho$ and
pressure $p$ ($T^{\mu}_{\ \nu}=diag(-\rho,p,p,p)$), then
Eq.~(\ref{rastal}) leads to

\begin{eqnarray}\label{rastal2}
\dot{\rho}+3H(\rho+p)=-\lambda\dot{R},
\end{eqnarray}

\noindent where dot denotes derivative with respect to time. It is
also useful to remind here that for the flat FRW universe

\begin{eqnarray}\label{ri}
R=6\big[(\frac{\dot{a}}{a})^2+\frac{\ddot{a}}{a}\big].
\end{eqnarray}

\noindent Moreover, the use of Eq.~(\ref{r1}) yields \cite{hm}

\begin{eqnarray}\label{friedman1}
&&(12\gamma-3)H^2+6\gamma\dot{H}=-\frac{4\gamma-1}{6\gamma-1}8\pi G \rho,\\
&&(12\gamma-3)H^2+(6\gamma-2)
\dot{H}=\frac{4\gamma-1}{6\gamma-1}8\pi G p,\nonumber
\end{eqnarray}

\noindent The evolution of density perturbation in this model has
been studied in Refs. \cite{prd,rascos1,obs1}. It has been shown
that the story is the same as those of the standard cosmology at
the background and linear perturbation level \cite{prd}. Finally,
one can use Eqs.~(\ref{friedman1}) to get \cite{hm}

\begin{equation}\label{rey}
\dot{H}=-\frac{k}{2}(\rho+p).
\end{equation}

\noindent From the standpoint of tensor calculus, Eq.~(\ref{r1})
is a solution for Eq.~(\ref{rastal}) leading to the above results.
But, does thermodynamics lead to the same solutions? Indeed, since
entropy is the backbone of thermodynamic approach, it is expected
that Eq.~(\ref{r1}) and thus the above results are available only
whenever the horizon entropy meets Eq.~(\ref{entf}). In order to
find the Friedmann equations corresponding on Eq.~(\ref{rastal})
from the thermodynamics point of view, we consider the general
form of entropy as $S_A=S(\frac{2\pi A}{k})$. Additionally, one
can use

\begin{eqnarray}\label{esv}
\delta Q^m = (T^b_a\partial_b \tilde{r} + W\partial_a
\tilde{r})dx^a,
\end{eqnarray}

\noindent to evaluate the energy flux crossing the apparent
horizon \cite{Cai2,CaiKim}. Here, we also focus on an
energy-momentum source as
$T_{\mu}^{\nu}=\textmd{diag}(-\rho,p,p,p)$ which yields
$W=\frac{\rho-p}{2}$ for the work density. Finally, we see
\cite{hm,plb1c}

\begin{eqnarray}\label{ufl3}
\delta Q^m=-A(\rho+p)dt.
\end{eqnarray}

\noindent Now, applying the Clausius relation ($TdS_A=-\delta
Q^m$) to horizon \cite{abe}, and using the Cai-Kim temperature
($T=\frac{H}{2\pi}$) \cite{CaiKimt}, one can easily find

\begin{equation}\label{rey1}
\dot{H}=-\frac{\pi}{S^{\prime}}(\rho+p),
\end{equation}

\noindent where $S^\prime=\frac{dS_A}{dA}$ \cite{pwang}. Now,
combining this result with Eq.~(\ref{rastal2}), we obtain

\begin{equation}\label{fdif}
S^{\prime}dH^2=\frac{2\pi}{3}d(\rho+\lambda R),
\end{equation}

\noindent as the differential form of the first Friedmann
equation. The result of this equation can be combined with
Eq.~(\ref{rey1}) to get the second Friedmann equation.

Now, if the system entropy meets Eq.~(\ref{entf}), then
Eq.~(\ref{rey1}) leads to $2\dot{H}=-k(\rho+p)$. It is also easy
to insert Eq.~(\ref{entf}) into Eq.~(\ref{fdif}) to get

\begin{eqnarray}\label{friedman1f}
(12\gamma-3)H^2+6\gamma\dot{H}+\mathcal{C}=-\frac{4\gamma-1}{6\gamma-1}8\pi
G \rho,
\end{eqnarray}

\noindent where, $\mathcal{C}$ is the integration constant of
Eq.~(\ref{fdif}). Now, adding and subtracting $2\dot{H}$ from the
LHS this equation, and using the $2\dot{H}=-k(\rho+p)$ relation,
we can reach at

\begin{eqnarray}\label{friedman1ff}
(12\gamma-3)H^2+(6\gamma-2)\dot{H}+\mathcal{C}=\frac{4\gamma-1}{6\gamma-1}8\pi
G p.
\end{eqnarray}

\noindent It is apparent that, for the $\mathcal{C}=0$ case, the
original Friedmann equations in the Rastall framework are
reproduced~(\ref{friedman1f}). Indeed, since divergence of
$\mathcal{C}g_{\mu\nu}$ is zero, one may add the
$\mathcal{C}g_{\mu\nu}$ term to the RHS of Rastall field
equations~(\ref{r1}) to directly get Eq.~(\ref{friedman1f})
instead of Eq.~(\ref{friedman1}). Anyway, we know that this term
represents an unusual fluid in the context of ordinary physics
leading to dark energy concept and thus its problems.

Now, Let us consider a flat FRW universe filled by a fluid with
constant state parameter defined as $w=\frac{p}{\rho}$. In this
manner, calculations lead to \cite{hm}

\begin{eqnarray}\label{den}
\rho=\rho_0
a^{\frac{-3(1+\omega)(4\gamma-1)}{3\gamma(1+\omega)-1}},
\end{eqnarray}

\noindent where $\rho_0$ is constant, for the energy density
profile. For a universe filled by a pressureless component, where
$\rho=\rho_0 a^{\frac{-3(4\gamma-1)}{3\gamma-1}}$, combining this
equation with Eqs.~(\ref{friedman1f}), one reaches at

\begin{eqnarray}\label{rey3}
\dot{H}=-\frac{\rho_0(4\gamma-1)4\pi
G}{6\gamma-1}a^{\frac{-3(4\gamma-1)}{3\gamma-1}},
\end{eqnarray}

\noindent and

\begin{eqnarray}\label{fr2f}
H^2=\frac{\rho_0(3\gamma-1)8\pi
G}{3(6\gamma-1)}a^{\frac{-3(4\gamma-1)}{3\gamma-1}}+\mathcal{C}.
\end{eqnarray}

Now, it is natural expectation that the matter density should be
diluted during the expansion of universe. This limits us to the
Rastall theories with $\gamma<\frac{1}{4}$ and
$\frac{1}{3}<\gamma$. Applying this result to Eq.~(\ref{rey3}), we
find out that, at the long run limit ($a(t)\gg1$), we have
$\dot{H}\rightarrow0$, and therefore, universe may experience an
accelerating phase. In this situation, Eq.~(\ref{fr2f}) implies
$H=H_0\approx \mathcal{C}$ meaning that a non-minimal coupling
between geometry and energy-momentum source cannot describe the
current accelerating phase of universe in the Rastall framework.
In fact, as it is apparent, we should have $\mathcal{C}\neq0$ to
get $H_0\neq0$, and thus, an accelerating universe. Therefore, the
same as the standard Friedmann equations, a dark energy-like
source is needed to model the accelerating universe in the Rastall
theory, a result in agreement with recent study by Batista et al.
\cite{prd}.

In summary, our thermodynamic based study shows the dark energy
problem is also valid in this formalism \cite{prd}, unless one
generalizes the Rastall theory in a suitable manner \cite{genras}.

\section{R\'{e}nyi entropy and Friedmann equations}

Recently, R\'{e}nyi entropy has been used in order to study the
effects of probable non-extensive aspects of spacetime which led
to interesting results in both cosmological and gravitational
setups \cite{rey1,rey2,EPJC,MSK,rey03,rey04,rey05,rey06}. For a
non-extensive system including $W$ discrete states, the R\'{e}nyi
entropy is defined as \cite{reyo1}

\begin{eqnarray}\label{reyn02}
\mathcal{S}=\frac{1}{1-q}\ln\sum_{i=1}^{W} P_i^q,
\end{eqnarray}

\noindent in which $P_i$ and $q$ denote the probability of
$i^{th}$ state and the non-extensive parameter, respectively.
Moreover, the Tsallis entropy of this system is as follows
\cite{tsallis}

\begin{eqnarray}\label{reyn01}
S_T=\frac{1}{1-q}\sum_{i=1}^{W}(P_i^q-P_i).
\end{eqnarray}

\noindent The linear relation between entropy and area is the key
point of the Bekenstein-Hawking entropy ($S_B\sim A$), which can
also be obtained from the Tsallis' non-additive entropy definition
\cite{rn3}. As it is apparent from Eq.~(\ref{entf}), the
functionality of $S_A^R$ with respect to the horizon area ($A$) is
the same as that of the Bekenstein-Hawking entropy ($S_A^R\sim A$)
meaning that $S_A^R$ may be considered as a special case of
Eq.~(\ref{reyn01}). More detailed studies on gravitational and
cosmological implications of Tsallis entropy~(\ref{reyn01}) can be
found in Refs.
\cite{rn3,rsal1,rsal2,rsal,anp,ijtp,cite5,Kom1,Kom2,Kom3} and
references therein. Eq.~(\ref{reyn01}) can be combined with
Eq.~(\ref{reyn02}) to show that

\begin{eqnarray}\label{reyn1}
\mathcal{S}=\frac{1}{\delta}\ln(1+\delta S_T),
\end{eqnarray}

\noindent where $\delta\equiv1-q$, and we used $\sum_{i=1}^{W}
P_i=1$ to obtain this equation \cite{rey1,MSK}. It has frequently
been argued that the Bekenstein-Hawking entropy is not an
extensive entropy
\cite{rn1,rn2,rn4,rn5,rn6,rn7,rn8,rn9,rn10,rn11,rn12,rn13,rn14,rn15,rn16,rey2},
and in fact, the Bekenstein-Hawking entropy can be considered as a
proper candidate for $S_T$ in gravitational and cosmological
setups \cite{rey1,rey2,EPJC,MSK}, a choice in full agreement with
Ref. \cite{rn3}. Here, following the above arguments and recent
studies \cite{rn3,rey1,rey2,EPJC,MSK}, we use Eq.~(\ref{entf}) as
the Tsallis entropy candidate in our calculations leading to

\begin{eqnarray}\label{renyi}
\mathcal{S}_A&=&\frac{1}{\delta}\ln(1+\delta
S_A^R)=\frac{1}{\delta}\ln(1+\frac{2\pi\delta}{k}A),\\
\label{renyi1}\mathcal{S}_A^\prime&=&\frac{d\mathcal{S}_A}{dA}=\frac{2\pi
H^2}{k[H^2+\Delta]},
\end{eqnarray}

\noindent for the R\'{e}nyi entropy of horizon ($\mathcal{S}_A$)
and its derivative with respect to $A$ ($\mathcal{S}_A^\prime$),
respectively. Here,
$\Delta\equiv\frac{(6\gamma-1)\delta\pi}{(4\gamma-1)G}$, and it is
easy to check that whenever the non-extensive features of system
approach zero (or equally $\delta\rightarrow0$), the
$\mathcal{S}_A$ relation recovers Eq.~(\ref{entf}). Now, inserting
Eq.~(\ref{renyi1}) into Eqs.~(\ref{rey1}) and~(\ref{fdif}), one
can obtain

\begin{equation}\label{reyf}
\dot{H}=-\frac{k[H^2+\Delta]}{2H^2}(\rho+p),
\end{equation}

\noindent and

\begin{equation}
H^2-\Delta\ln(\Delta+H^2)+C_1=\frac{k}{3}\rho+\frac{\gamma}{3}R,
\end{equation}

\noindent where $C_1$ is the integration constant, respectively.
Defining a new constant $C=C_1-\Delta\ln\Delta$, one can rewrite
the last equation as

\begin{equation}\label{fff}
H^2-\Delta\ln(1+\frac{H^2}{\Delta})+C=\frac{k}{3}\rho+\frac{\gamma}{3}R.
\end{equation}

Since $H=\frac{\dot{a}}{{a}}$, Eq.~(\ref{ri}) can be rewritten as

\begin{equation}\label{ri1}
R=6(\dot{H}+2H^2).
\end{equation}

\noindent Finally, combining this equation with Eq.~(\ref{fff}),
and inserting the result into Eq.~(\ref{reyf}), we find

\begin{equation}\label{fff10}
H^2=\frac{k}{3}(\rho+\rho_e^\gamma),
\end{equation}

\noindent and

\begin{eqnarray}\label{fff1}
&&H^2+\frac{2}{3}\dot{H}=-\frac{k}{3}(p+p_e^\gamma),
\end{eqnarray}

\noindent for the first and second Friedmann equations in our
model, respectively. In fact, one should combine
Eqs.~(\ref{fff10}) and~(\ref{reyf}) with each other, and then add
and subtract the $\frac{2}{3}\dot{H}$ term to the result to find
the last equation. In the above equations,

\begin{eqnarray}\label{eff1}
\rho_e^\gamma&=&\frac{3}{k}\bigg(4\gamma H^2+\Delta\ln(1+\frac{H^2}{\Delta})+2\gamma\dot{H}-C\bigg),\nonumber\\
p_e^\gamma&=&-\frac{3}{k}\bigg(4\gamma
H^2+\Delta\ln(1+\frac{H^2}{\Delta})\\&+&2\dot{H}(\gamma+\frac{1}{3(\frac{H^2}{\Delta}+1)})-C\bigg),\nonumber
\end{eqnarray}

\noindent denote effective energy density and effective pressure
in this framework, respectively. It is easy to see that, at the
$\delta\rightarrow0$ limit (or equally $\Delta\rightarrow0$), the
results of Rastall theory are recovered. Additionally, we have
$p_e\rightarrow-\rho_e$ whenever $\dot{H}\rightarrow0$. Now, since
$\frac{\ddot{a}}{a}=H^2+\dot{H}$, one can rewrite Eq.~(\ref{fff1})
as

\begin{eqnarray}\label{refer1}
&&2\frac{\ddot{a}}{a}+H^2=-k(p+p_e^\gamma),
\end{eqnarray}

\noindent where its LHS has the same form as that of the standard
Friedmann equation \cite{roos}. Besides, simple calculations reach

\begin{eqnarray}\label{refer2}
&&\frac{\ddot{a}}{a}=-\frac{k}{6}[\rho+\rho_e^\gamma+3(p+p_e^\gamma)],
\end{eqnarray}

\noindent for the acceleration equation. Therefore, we deal with
two fluids. The first fluid is the ordinary energy-momentum tensor
$T_\mu^\nu$ corresponding to the real fluid with energy density
$\rho$ and pressure $p$. The second fluid, which has geometrical
origin, is called the effective energy-momentum tensor, and it is
defined as

\begin{eqnarray}\label{efemt}
\Theta_\mu^\nu=diag(-\rho_e^\gamma,p_e^\gamma,p_e^\gamma,p_e^\gamma).
\end{eqnarray}

\noindent In fact, the obtained effective fluid consists of two
parts: $i)$ the non-extensive aspects of spacetime, and $ii)$ the
non-minimal coupling between geometry and matter fields which
follows the Rastall hypothesis. Therefore, the
$\gamma\rightarrow0$ limit of $\Theta_\mu^\nu$ only includes the
non-extensive effects. We show it by
$\mathcal{T}_\mu^\nu=diag(-\rho_e,p_e,p_e,p_e)$ in which

\begin{eqnarray}\label{eff}
&&\rho_e=\frac{3}{8\pi G}\bigg(\Delta_{(\gamma=0)}\ln(1+\frac{H^2}{\Delta_{(\gamma=0)}})-C_{(\gamma=0)}\bigg),\nonumber\\
&&p_e=-\rho_e-\frac{3}{8\pi
G}\bigg(\frac{2\dot{H}}{3(\frac{H^2}{\Delta_{(\gamma=0)}}+1)}\bigg),
\end{eqnarray}

\noindent where $\Delta_{(\gamma=0)}=\frac{\delta\pi}{G}$ and
$C_{(\gamma=0)}=C_1-\Delta_{(\gamma=0)}\ln\Delta_{(\gamma=0)}$. In
fact, it is a geometrical fluid originated from the non-extensive
aspects of spacetime, and recovers the ordinary cosmological
constant model of dark energy at the appropriate limit of
$\gamma\rightarrow0$. It is also easy to check that this source
satisfies the conservation law i.e.

\begin{eqnarray}\label{emcl3}
\dot{\rho}_e+3H(\rho_e+p_e)=0,
\end{eqnarray}

\noindent Therefore, $\Theta_\mu^\nu$ acts as a time-varying dark
energy model which satisfies the conservation law only for
$\gamma=0$.

Bearing the Bianchi identity in mind, since the LHS of
Eqs.~(\ref{fff10}) and~(\ref{refer1}) are compatible with the
Einstein tensor, we should have
$(\Theta_\mu^\nu+T_\mu^\nu)^{;\mu}=0$ leading to

\begin{eqnarray}\label{emclf}
\dot{\rho}+3H(\rho+p)=-[\dot{\rho}_e^\gamma+3H(\rho_e^\gamma+p_e^\gamma)],
\end{eqnarray}

\noindent and thus

\begin{eqnarray}\label{emcl1f}
\lambda\dot{R}=\dot{\rho}_e^\gamma+3H(\rho_e^\gamma+p_e^\gamma).
\end{eqnarray}

\noindent In fact, the above results would also be obtained by
writing Einstein field equations as
$G_{\mu\nu}=k(\Theta_{\mu\nu}+T_{\mu\nu})$. In addition,
Eqs.~(\ref{emcl3}) and~(\ref{emcl1f}) tell us that there is no
energy flux between geometry and matter fields at the appropriate
limit of $\lambda\rightarrow0$ (or equally $\gamma\rightarrow0$),
a result in full agreement with Eq.~(\ref{rastal}). Indeed,
although $\Theta_\mu^\nu\rightarrow\mathcal{T}_\mu^\nu$ at the
$\gamma\rightarrow0$ limit, since $\mathcal{T}_\mu^\nu$ is a
divergence-less tensor~(\ref{emcl3}), we have $\lambda\dot{R}=0$
and thus the ordinary energy-momentum conservation law is met by
the $T_\mu^\nu$ source. Applying the $\gamma\rightarrow0$ and
$\Delta\rightarrow0$ limits to the above equations, one can easily
reach at the standard Friedmann equations compatible with the
Bekenstein-Hawking entropy of horizon \cite{ijtp,md,em}.
Therefore, the $\gamma\rightarrow0$ limit helps us in obtaining
the modification of considering R\'{e}nyi entropy to the standard
Friedmann equations as

\begin{eqnarray}\label{mfe}
&&H^2=\frac{8\pi G}{3}(\rho+\rho_e),\\
&&H^2+\frac{2}{3}\dot{H}=\frac{-8\pi G}{3}(p+p_e),\nonumber
\end{eqnarray}

\noindent where $\rho_e$ and $p_e$ follow Eq.~(\ref{eff}). It is
worth to mention that, independent of the values of $C$ and
$\Delta$, we have $p_e\rightarrow-\rho_e$ whenever
$\dot{H}\rightarrow0$. As a check, one can also insert $\gamma=0$
in Eqs.~(\ref{k}) and~(\ref{eff1}) to get these results. In this
manner, the acceleration equation is

\begin{eqnarray}\label{refer4}
&&\frac{\ddot{a}}{a}=-\frac{4\pi G}{3}[\rho+\rho_e+3(p+p_e)],
\end{eqnarray}

\noindent and the second line of Eq.~(\ref{mfe}) can also be
written as

\begin{eqnarray}\label{refer3}
&&2\frac{\ddot{a}}{a}+H^2=-8\pi G(p+p_e),
\end{eqnarray}

\noindent where its LHS is in the form of the standard Friedmann
equation \cite{roos}. Bearing Eq.~(\ref{emcl3}) as well as the
argument after Eq.~(\ref{emcl1f}) in mind, it is apparent that,
for $\gamma=0$, the $T_{\mu\nu}$ source respects the continuity
equation i.e.

\begin{eqnarray}\label{emcl2}
\dot{\rho}+3H(\rho+p)=0.
\end{eqnarray}

Although the same as Refs.~\cite{EPJC,MSK}, we used the R\'{e}nyi
entropy to get the modified Friedmann equations, these equations
differ from those of recent studies \cite{EPJC,MSK}. It has three
reasons. $\textmd{i}$) While the entropy expression appears in
acceleration equations obtained in Refs.~\cite{EPJC,MSK}, which
use the Padmanabhan approach, its derivative is the backbone of
getting the acceleration equation in our model based on applying
thermodynamics laws to horizon (see Eq.~(\ref{rey1})).
$\textmd{ii}$) The Komar mass definition has been used by authors
in Refs. \cite{EPJC,MSK} only for the $T_{\mu\nu}$ source. As we
saw, our results are also obtainable if one writes the Einstein
field equations as $G_{\mu\nu}=k(\Theta_{\mu\nu}+T_{\mu\nu})$
meaning that the Komar mass should be written for the modified
energy-momentum tensor $\Theta_{\mu\nu}+T_{\mu\nu}$ instead of
$T_{\mu\nu}$. $\textmd{iii}$) In Ref.~\cite{EPJC}, the
non-extensive features of spacetime has been introduced as an
origin for a time-varying dark energy ($\Lambda(t)$) which
interacts with matter fields and does not meet Eq.~(\ref{emcl3}).
This is while time-varying dark energy candidate of our model
interacts with matter fields only in the Rastall way, and meets
Eq.~(\ref{emcl3}) in the absence of the Rastall hypothesis
($\gamma=0$).

\section{A Universe filled by a pressureless fluid}

In order to study a universe filled by a pressureless source, we
insert $\rho=\rho_0 a^{\frac{-3(4\gamma-1)}{3\gamma-1}}$ into
Eq.~(\ref{reyf}) and use Eq.~(\ref{fff10}) to reach at

\begin{eqnarray}\label{fr2}
\frac{H^2(1-4\gamma)-\Delta\ln(1+\frac{H^2}{\Delta})+C}{1-3\gamma(1+\frac{\Delta}{H^2})}=\xi
a^{\frac{-3(4\gamma-1)}{3\gamma-1}},
\end{eqnarray}

\noindent where $\xi\equiv\frac{\rho_0(4\gamma-1)8\pi
G}{3(6\gamma-1)}$. It is clear that, for a Rastall theory of
$\gamma<\frac{1}{4}$ or $\frac{1}{3}<\gamma$, the RHS of this
equation and thus its LHS are vanished at long run limit ($a\gg1$)
meaning that $H\rightarrow constant\equiv H_0$. Now, $H_0$ can be
evaluated from

\begin{eqnarray}\label{H}
\frac{H_0^2}{\Delta}=\frac{\ln(1+\frac{H_0^2}{\Delta})}{1-4\gamma}+C_2,
\end{eqnarray}

\noindent where $C_2\equiv\frac{C}{(4\gamma-1)\Delta}$ is a
constant. It is also apparent that, depending on the value of
$\gamma$, this equation may be solvable even if we have $C_2=0$ or
$C_1=0$. Additionally, since $\Delta$ is unknown parameter, this
equation helps us in finding its possible values as a function of
$H_0$. One can also use this result in order to apply the $a\gg1$
limit to Eq.~(\ref{reyf}) to see that $\dot{H}\rightarrow0$
whenever $\gamma$ meets either $\gamma<\frac{1}{4}$ or
$\frac{1}{3}<\gamma$. In summary, based on our results, the dark
energy problem in Rastall theory can be overcame by considering
the probable non-extensive features of spacetime.

Now, we consider a universe filled by a pressureless source
($p=0$) whenever $\gamma=0$. In this manner, both of
Eqs.~(\ref{den}) and~(\ref{emcl2}) lead to $\rho=\rho_0 a^{-3}$
for energy density. Inserting $\gamma=0$ into Eqs.~(\ref{mfe}),
and following the recipe which led to Eqs.~(\ref{fr2})
and~(\ref{H}), we reach at

\begin{eqnarray}\label{frff}
H^2-\Delta_{(\gamma=0)}\ln(1+\frac{H^2}{\Delta_{(\gamma=0)}})+C_{(\gamma=0)}=\chi
a^{-3},
\end{eqnarray}

\noindent and

\begin{eqnarray}\label{Hf}
\frac{H_0^2}{\Delta_{(\gamma=0)}}=\ln(1+\frac{H_0^2}{\Delta_{(\gamma=0)}})-C_3,
\end{eqnarray}

\noindent where $\chi\equiv\frac{8\pi G\rho_0}{3}$, and
$C_3\equiv\frac{C_{(\gamma=0)}}{\Delta_{(\gamma=0)}}$. It also
means that, whenever the divergence of $T_\mu^\nu$ is zero, the
probable non-extensive features of spacetime, which behave as a
conserved fluid, can be considered as the nature of current
accelerating phase of universe if $\Delta_{(\gamma=0)}$ and $H_0$
meet the above equation. It is worth to note here that this
equation is solvable even if $C_1=0$. This result (the $\gamma=0$
case) is in agreement with our previous results, where we found
out the $\gamma$ parameter should either meet $\gamma<\frac{1}{4}$
or $\frac{1}{3}<\gamma$.

Bearing the results addressed after Eqs.~(\ref{fr2f})
and~(\ref{H}) in mind, it is worth to mention that from the view
point of dynamics, the $\gamma<\frac{1}{4}$ and
$\frac{1}{3}<\gamma$ intervals are permissible for $\gamma$. On
the other hand, thermodynamic considerations (the results of
Eq.~(\ref{entf})), insist only the $\gamma<\frac{1}{6}$ and
$\frac{1}{4}<\gamma$ intervals are admissible. Comparing these
results with each other, one can easily find that
$\gamma<\frac{1}{6}$ and $\frac{1}{3}<\gamma$ are common
intervals. This means that the values of $\gamma$ obtained from
observations are allowed, if they be within in these ranges.

\section{Observational Constraints}

In what follows, let us discuss the observational constraints on
the scenarios presented above. In order to constrains the free
parameters of the models we use: the Union $2.1$ sample
\cite{2012ApJ...746...85S}, which contains $580$ Supernovae type
Ia (SNIa) in the redshift range $0.015 \leq z \leq 1.41$,  36
Observational Hubble Data ($H(z)$) in the range ($0.0708 \leq z
\leq 2.36$) compiled in \cite{2015arXiv150702517M} and the Baryon
Acoustic Oscillations (BAO) distance measurements at different
redshift, in order to diminish the degeneracy between the free
parameters.

\subsection{Supernovae type Ia}

In order to study the constraints applied to a cosmological model
by the SNIa data, one can use the distance modulus $\mu (z)$
defined as

\begin{equation}
\mu_{th}(z) = 5 \log_{10} D_L(z) +\mu_0,
\end{equation}

\noindent where $\mu_0=42.38-5log_{10}h$. Moreover, $h=H_0/100\
km\cdot s^{-1}\cdot Mp\ c^{-1}$ is the dimensionless Hubble
parameter, and $D_L(z)$ is the luminosity distance calculated as

\begin{equation}
D_L(z)=\frac{c(1+z)}{H_0} \int^z_0 \frac{dz'}{E(z')},
\end{equation}

\noindent in the flat FRW universe. Here, $E^2(z)=H^2(z)/H^2_0$,
and thus, using Eqs.~(\ref{fff10}) and~(\ref{mfe}), we can easily
reach

\begin{eqnarray}\label{50}
 E^2(z) &=& \Omega_m (1+z)^3 + 4 \gamma  E^2(z) +  \\
&& \frac{\Delta}{H^2_0}\ln \left( 1 + \frac{H^2_0}{\Delta}  E^2(z) \right) + 2 \gamma \frac{\dot{H}}{H^2_0} + \Omega_C \nonumber
\end{eqnarray}

\noindent whenever $\gamma\neq0$ and $\Omega_{C} = - C/H_0^2$
(Model $\textmd{I}$), and

\begin{eqnarray}\label{51}
E^2(z) &=& \Omega_m (1+z)^3  +  \\
&& \frac{\Delta_{(\gamma=0)}}{H^2_0}\ln \left( 1 +
\frac{H^2_0}{\Delta_{(\gamma=0)}}  E^2(z) \right)  +
\Omega_{C_{(\gamma=0)}}, \nonumber
\end{eqnarray}

\noindent while $\gamma=0$ and $\Omega_{C_{(\gamma=0)}} = -
C_{(\gamma=0)}/H_0^2$ (Model $\textmd{II}$), respectively. In the
above results, $ \Omega_m = \rho_{m0}/\rho_{cr}$ and $\rho_{cr}$
is the critical density defined as $\rho_{cr}=3H_0^2/8\pi G$. Now,
applying the $E^2(z=0)=1$ and $\dot{H}(z=0)=0$ conditions (the
usual normalization conditions at $z=0$) to Eqs.~(\ref{50})
and~(\ref{51}), we get

\begin{equation}
\Omega_{C} = 1 - \Omega_{m0} - 4 \gamma - \frac{\Delta}{H_0^2} \ln \left( 1 + \frac{H_0^2}{\Delta}\right),
\end{equation}

\noindent and

\begin{equation}
\Omega_{C_{(\gamma=0)}} = 1 - \Omega_{m0} -
\frac{\Delta_{(\gamma=0)}}{H_0^2} \ln \left( 1 +
\frac{H_0^2}{\Delta_{(\gamma=0)}}\right),
\end{equation}

\noindent in Model $\textmd{I}$ and Model $\textmd{II}$,
respectively. Therefore, Model $\textmd{I}$ has three free
parameters as $\left\lbrace \Omega_{m0}, \Delta,
\gamma\right\rbrace $, and Model $\textmd{II}$ has two free
parameters including $\left\lbrace \Omega_{m0},
\Delta_{(\gamma=0)}\right\rbrace$. It is worthwhile to remind here
that if $\Delta \rightarrow 0$ and $\gamma \rightarrow 0$, then
the $\Lambda CDM$ model is recovered for $C = C_1 \equiv \Lambda$.

Observational constraints on cosmological model can be obtained by
minimizing  $\chi^2$ given by \cite{metodo,2005PhRvD..72l3519N}

\begin{equation}
\chi_{SNIa}^2 = \textsf{A} - \frac{\textsf{B}^2}{\textsf{C}},
\end{equation}

\noindent where

\begin{eqnarray}
&&\textsf{A} = \sum_{i=1}^{580}\frac{
[\mu_{th}(z_{i},p_i)-\mu_{obs}(z_{i})]^2 }{\sigma_{\mu_i}^2},\nonumber\\
&&\textsf{B} = \sum_{i=1}^{580}\frac{
\mu_{th}(z_{i},p_i)-\mu_{obs}(z_{i}) }{\sigma_{\mu_i}^2},\nonumber\\
&&\textsf{C} = \sum_{i=1}^{580}\frac{1}{\sigma_{\mu_i}^2},
\end{eqnarray}

\noindent and we have marginalized over the nuisance parameter
$\mu_0$ and $\mu_{obs}$.

\subsection{Baryon Acoustic Oscillations (BAO)}

The expanding spherical wave of baryonic perturbations, which
comes from acoustic oscillations at recombination and co-moving
scale of about $150Mpc$, helps us in identifying the peak of large
scale correlation function measured from SDSS (Sloan Digital Sky
Survey). It is worth to note that the BAO scale depends on
$\textmd{i}$) the scale of sound horizon at recombination,
$\textmd{ii}$) the transverse and radial scales at the mean
redshift of galaxies in the survey. In order to obtain the
corresponding constraints on the cosmological models, we begin
with $\chi^2$ for the WiggleZ BAO data \cite{2011MNRAS.415.2892B}
given as

\begin{equation}
\chi^2_{\scriptscriptstyle WiggleZ} =
(\bar{A}_{obs}-\bar{A}_{th})C_{\scriptscriptstyle
WiggleZ}^{-1}(\bar{A}_{obs}-\bar{A}_{th})^T,
\end{equation}

\noindent where $\bar{A}_{obs} = (0.447, 0.442, 0.424)$ is data
vector at $z=(0.44,0.60,0.73)$, $T$ denotes the ordinary
transpose, and $\bar{A}_{th}(z,p_i)$ is \cite{2005ApJ...633..560E}

\begin{equation}
\bar{A}_{th}=D_V(z) \frac{\sqrt{\Omega_m H_0^2}}{cz},
\end{equation}

\noindent in which

\begin{equation}
D_V(z) = \frac{1}{H_0}\left[ (1+z)^2 D_A (z)^2
\frac{cz}{E(z)}\right]^{1/3},
 \end{equation}

\noindent is the distance scale. Here, $D_A(z)$ denotes the
angular diameter distance defined as
$D_A(z)=\frac{D_L(z)}{(1+z)^2}$. Additionally,
$C_{\scriptscriptstyle WiggleZ}^{-1}$ is the inverse covariance
matrix for the WiggleZ data set given by

\begin{equation}
C_{\scriptscriptstyle WiggleZ}^{-1} = \left(
\begin{array}{ccc}
1040.3 & -807.5   & 336.8    \\
-807.5  & 3720.3  & -1551.9 \\
336.8   & -1551.9 & 2914.9
\end{array}\right).
\end{equation}

For the SDSS DR7 BAO distance measurements, $\chi^2$ can similarly
be expressed as \cite{2010MNRAS.401.2148P}

\begin{equation}
\chi^2_{\scriptscriptstyle SDSS} =
(\bar{d}_{obs}-\bar{d}_{th})C_{\scriptscriptstyle
SDSS}^{-1}(\bar{d}_{obs}-\bar{d}_{th})^T,
 \label{eq3:3.19}
\end{equation}

\noindent where $\bar{d}_{obs} = (0.1905,0.1097)$ is the data
points at $z=0.2$ and $z=0.35$. $\bar{d}_{th}(z_d,p_i)$ is also
defined as

\begin{equation}
\bar{d}_{th} = \frac{r_s(z_d)}{D_V(z)},
 \label{eq3:3.20}
\end{equation}

\noindent in which $r_s(z)$ is the radius of co-moving sound
horizon given by

\begin{equation}
 r_s(z) = c \int_z^\infty \frac{c_s(z')}{H(z')}dz',
  \label{eq3:3.13}
 \end{equation}

\noindent and

\begin{equation}
c_s(z) = \frac{1}{\sqrt{3(1+\bar{R_b}/(1+z)}},
 \label{eq3:3.14}
\end{equation}

\noindent is the sound speed. Here, $\bar{R_b} = 31500
\Omega_{b}h^2(T_{CMB}/2.7\rm{K})^{-4}$ and $T_{CMB}$ = 2.726K.
$z_{drag}$ at the baryon drag epoch fitted with the formula,
proposed in \cite{1998ApJ...496..605E},

\begin{equation}
z_{drag} =
\frac{1291(\Omega_{m}h^2)^{0.251}}{1+0.659(\Omega_{m}h^2)^{0.828}}[1+b_1(\Omega_b
h^2)^{b_2}],
 \label{eq3:3.15}
\end{equation}

\noindent where

\begin{equation}
 b_1 = 0.313(\Omega_{m}h^2)^{-0.419}[1+0.607(\Omega_{m}h^2)^{0.674}]
 \end{equation}

\noindent and

\begin{equation}
 b_2 = 0.238(\Omega_{m}h^2)^{0.223}.
\end{equation}

\noindent Here,

\begin{equation}
C_{\scriptscriptstyle SDSS}^{-1} = \left(
\begin{array}{cc}
30124 & -17227\\
-17227 & 86977
\end{array}\right),
 \label{eq3:3.21}
\end{equation}

\noindent is the inverse covariance matrix for the SDSS data set.
Additionally, we use the Six Degree Field Galaxy Survey (6dF)
measurement \cite{bao1}, the Main Galaxy Sample of Data Release
$7$ of Sloan Digital Sky Survey (SDSS-MGS) \cite{bao3}, the LOWZ
and CMASS galaxy samples of the Baryon Oscillation Spectroscopic
Survey (BOSS-LOWZ) \cite{bao3} and the distribution of the
LymanForest in BOSS (BOSS - $Ly_{\alpha}$) \cite{bao4}. These
measurements and their corresponding effective redshifts ($z$) are
summarized in Table \ref{BaoDat}. Therefore, the total
$\chi^2_{BAO}$ includes $9$ data point (for all the BAO data sets)

\begin{table}[H]
\begin{center}
\begin{tabular}{ccccc}
\hline
\hline
Survey                           & z & Parameter & Measurement &  Reference \\

\hline

6dF                                & 0.106  &  $r_s/D_V$ & $ 0.336 \pm 0.015$     & \cite{bao1} \\

SDSS-MGS                   & 0.57    &  $r_s/D_V$  & $0.0732 \pm 0.0012$ &  \cite{bao3} \\

BOSS-LOWZ                 & 0.32    &  $D_V/r_s$  & $8.47 \pm 0.17$         & \cite{bao3} \\

BOSS - $Ly_{\alpha}$    & 2.36    &  $D_A/r_s$  &  $10.08 \pm 0.4$       & \cite{bao4} \\

\hline
\end{tabular}
\end{center}
  \caption{Baryon acoustic oscillations (BAO) data measurements used in our statistical analysis.}
      \label{BaoDat}
\end{table}

\begin{eqnarray}
\chi_{BAO}^{2} &=& \chi^{2}_{WiggleZ} + \chi^{2}_{SDSS} + \chi^{2}_{6dF} + \chi^{2}_{SDSS-MGS}\nonumber \\
 && + \chi^{2}_{BOSS-LOWZ} + \chi^{2}_{BOSS-Ly_{\alpha}}
\end{eqnarray}

\subsection{History of the Hubble parameter}

The differential evolution of early type passive galaxies provides
direct information about the Hubble parameter $H(z)$. We adopt
$36$ Observational Hubble Data (OHD) at different redshifts
($0.0708 \leq z \leq 2.36$) obtained from
\cite{2015arXiv150702517M}, where $26$ data are deduced from the
differential age method, and the remaining $10$ data belong to the
radial BAO method. Here, we use these data to constrain the
cosmological free parameters of the models under consideration.
The corresponding $\chi^2$ can be defined as \cite{metodo}

\begin{equation}
\chi_{H(z)}^2 (H_0,p_i) =\sum_{i=1}^{36}\frac{ [H_{obs} (z_i) - H_{th}(z_i,H_0,p_i)]^2 }{\sigma_{H}^2(z_i)},
\end{equation}

\noindent where $H_{th}(z_i,H_0,p_i)$ is the theoretical value of
the Hubble parameter at the redshift $z_i$. This equation can be
re-written as \cite{metodo}

\begin{equation}
\chi_{H(z)}^2 (H_0,p_i) = \textsf{A}_1 - \textsf{B}_1 +
\textsf{C}_1,
\end{equation}

\noindent in which

\begin{eqnarray}
&&\textsf{A}_1 = H_0^2 \sum_{i=1}^{36}\frac{
E^2(z_i,p_i)}{\sigma_i^2},\nonumber\\ &&\textsf{B}_1 =
2H_0\sum_{i=1}^{36} \frac{H_{obs} (z_i) E^2(z_i,p_i)}{\sigma_i^2},\nonumber\\
&&\textsf{C}_1 = \frac{H^2_{obs} (z_i) }{\sigma_i^2}.
\end{eqnarray}

\noindent The function $\chi_{H(z)}^2$ depends on the model
parameters. To marginalize over $H_0$, we assume that the
distribution of $H_0$ is a Gaussian function with standard
deviation width $\sigma_{H_0}$ and mean $\bar{H}_0$. Then, we
build the posterior likelihood function $\mathcal{L}_H(p)$ that
depends just on the free parameters $p_i$, as

\begin{equation}
\mathcal{L}_H(p_i) = \int \pi_H (H_0) exp \left[-\chi^2_H(H_0,p_i) \right] dH_0,
\end{equation}

\noindent where

\begin{equation}
\pi_H (H_0) = \frac{1}{\sqrt{2\pi}\sigma_{H_0}}  exp \left[ -\frac{1}{2} \left( \frac{H_0 - \bar{H}_0 }{\sigma_{H_0}}  \right)^2 \right] ,
\end{equation}

\noindent is a prior probability function widely used in the
literature. Finally, we minimize $\chi_{H(z)}^2 (p_i)=-2 \ln
\mathcal{L}_H(p_i)$ with respect to the free parameters $p_i$ to
obtain the best-fit parameters values.

\subsection{Statistic analysis and results}

Maximum likelihood $\mathcal{L}_{max}$, is the procedure of
finding the value of one or more parameters for a given statistic
which maximizes the known likelihood distribution. The maximum
likelihood estimate for the best fit parameters $p_{i}$ is

\begin{equation}
 \mathcal{L}_{max}(p_{i})= exp \left[ -\frac{1}{2}{\chi_{min}^{2}(p_{i})}
 \right],
  \label{eq4:4.1}
\end{equation}

\noindent and therefore, $\chi_{min}^{2}(p_{i})=-2 \ln
\mathcal{L}_{max}(p_{i})$ \cite{2010arXiv1012.3754A}. In order to
find the best values of the free parameters of the models, we
consider

\begin{equation}
\chi_{total}^{2}=\chi^{2}_{SNIa} +\chi^{2}_{BAO}+ \chi^{2}_{H(z)}.
\end{equation}

Moreover, the Fisher matrix is widely used in analyzing the
constraints of cosmological parameters from different
observational data sets
\cite{2009arXiv0901.0721A,2012JCAP...09..009W}. Having the best
fit $\chi_{min}^{2}(p_i, \sigma_i^2)$, the Fisher matrix can be
calculated as

\begin{equation}
F_{ij}=\frac{1}{2}\frac{\partial ^2 \chi_{min}^{2}}{\partial p_i \partial p_j},
\label{eq4:4.4}
\end{equation}

\noindent where $F_{ij}\big(\equiv F_{ij}(p_i, \sigma_i^2)\big)$
depends on the uncertainties $\sigma_i^2$ of the parameters $p_i$
for a given model. The inverse of the Fisher matrix also provides
an estimate of the covariance matrix through $\left[ C_{cov}
\right] = \left[ F \right]^{-1}$. Its diagonal elements are the
squares of uncertainties in each parameter marginalizing over the
others, while the off-diagonal terms yield the correlation
coefficients between parameters. The uncertainties obtained in the
propagation of errors are also given by $\sigma_i = \sqrt{Diag
\left[ C_{cov} \right]_{ij} }$. Note that the marginalized
uncertainty is always greater than (or at most equal to) the
non-marginalized one. In fact, marginalization cannot decrease the
error, and it has no effect, if all other parameters are
uncorrelated with it. Previously known uncertainties of
parameters, known as priors, can trivially be added to the
calculated Fisher matrix.

\begin{figure*}[!htbp]
\includegraphics[width=8cm,height=10cm]{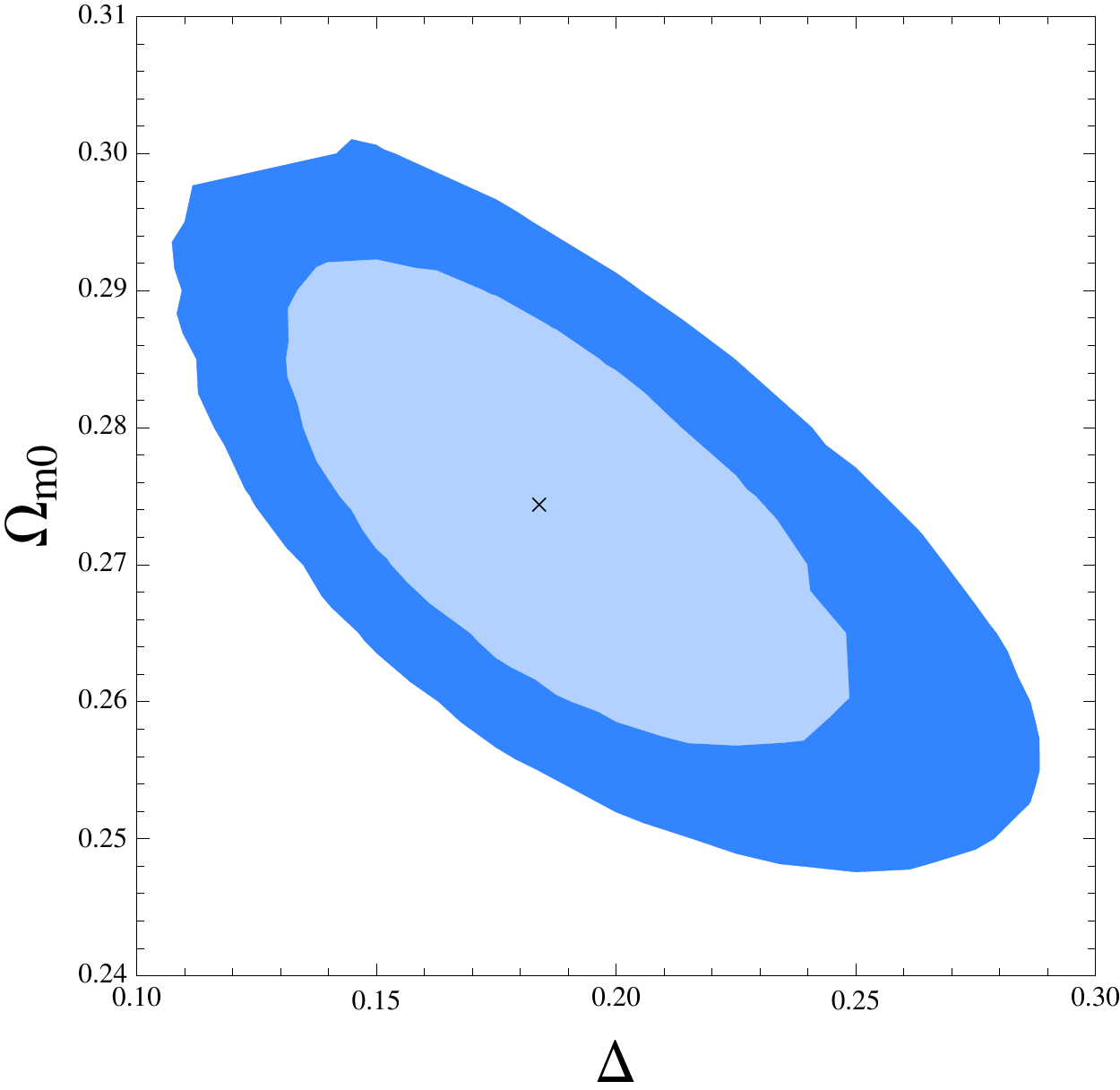}
\includegraphics[width=8cm,height=9.8cm]{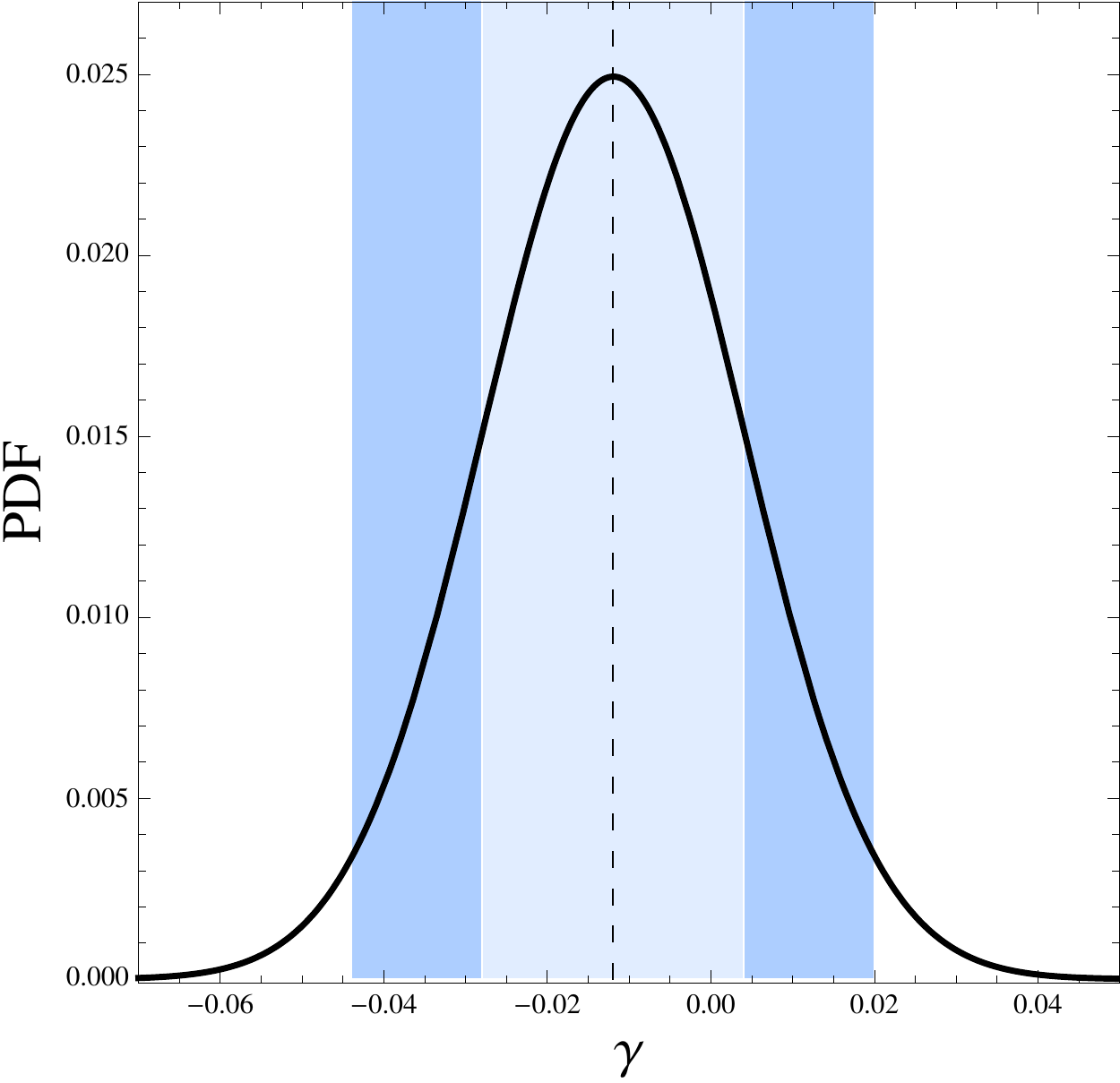}
\caption{Contour plots for the free parameter $\left\lbrace
\Delta- \Omega_{m0}\right\rbrace $ at $1\sigma$ and $2\sigma$ CL
for Model I, from the joint analysis $SNIa + BAO + H(z)$ (left
panel). Additionally, we present the corresponding marginalized
one-dimensional posterior distributions for $\gamma$ parameter
(right panel).}
   \label{Om_gam1}
\end{figure*}

\begin{figure}[!htbp]
\includegraphics[width=8cm,height=10cm]{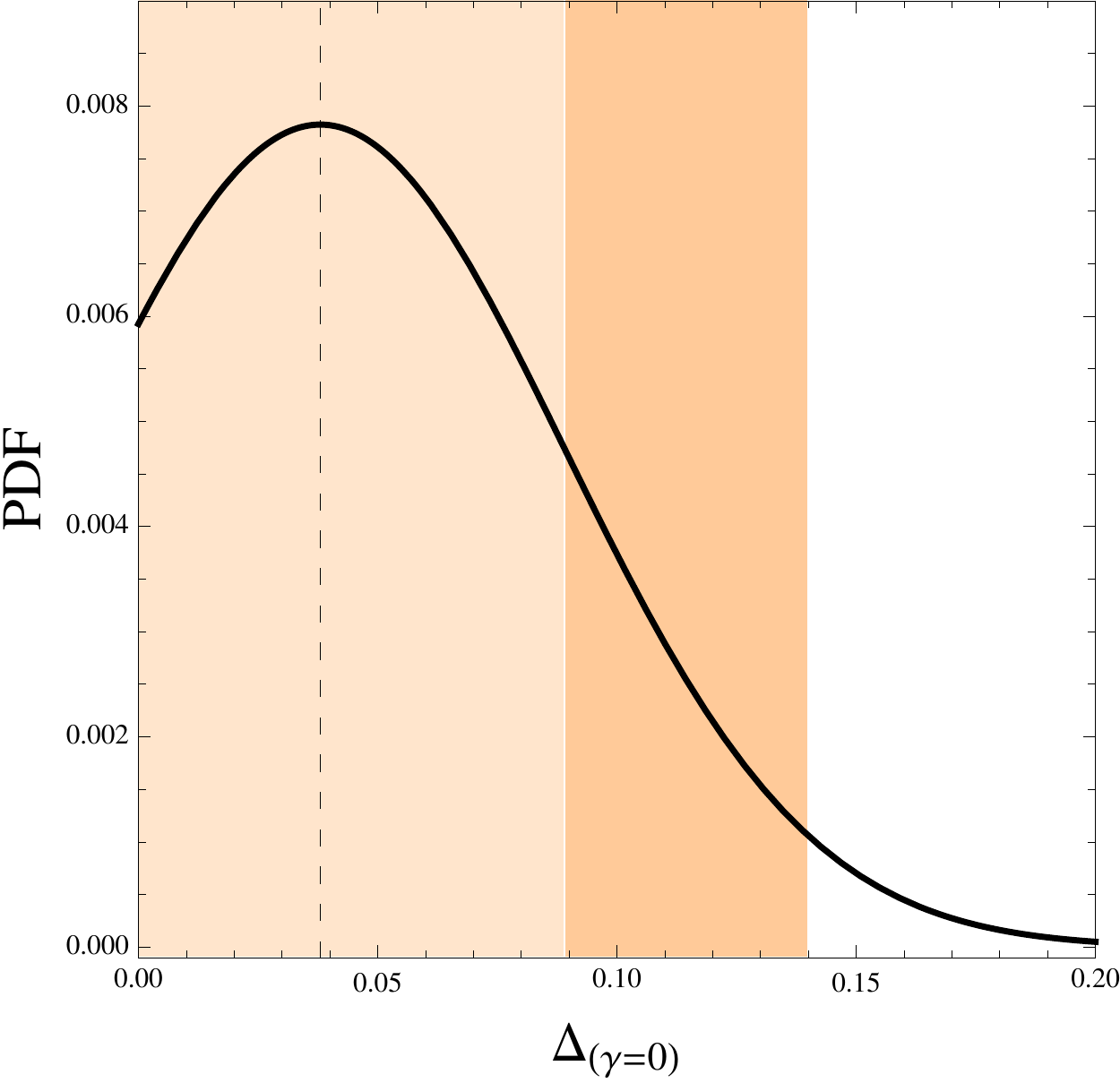}
\caption{Marginalized one-dimensional posterior distributions for
$\Delta_{(\gamma=0)}$ parameter at $1\sigma$ and $2\sigma$ CL for
Model II, from the joint analysis $SNIa + BAO + H(z)$.}
\label{Om_gam0}
\end{figure}

Table \ref{Om_del_gam} summarizes the main results of the
statistical analysis carried out by using the data sets SNIa, SNIa
+ BAO, and SNIa + BAO + $H(z)$ for two scenarios including
$\textmd{i}$) the R\'{e}nyi entropy is taking into account within
Rastall framework (Model I), and $\textmd{ii}$) the particular
case of $\gamma=0$ (Model II). The parameter $\Omega_{m0}$ takes
into account the content of cold dark matter plus baryons to the
present. It is useful to note that SNIa does not constrain
$\Delta$ very well, and in fact, results are improved by
introducing the other observational tests including $BAO$ and
$H(z)$. We can also see that the sign change of $\gamma$ does not
affect the main thermodynamic consideration obtained from
Eq.~(\ref{entf}). Indeed, since its obtained values meet the
$\gamma<\frac{1}{6}$ condition, entropy is always positive and
dynamics, i.e. the acceleration of the universe, is in agreement
with the observational data, an outcome in agreement with the
results of previous section. For Model $\textmd{I}$, the
likelihood contours arisen from the fitting analysis for the set
of free parameters ($\Delta$, $\Omega_{m0}$), and marginalized
one-dimensional posterior distributions for $\gamma$ (PDF),
considering the best fit values for used data sets $SNIa + BAO +
H(z)$, are presented in Fig.~\ref{Om_gam1}. Moreover,
Fig.~\ref{Om_gam0}. includes marginalized one-dimensional
posterior distributions for $\Delta_{(\gamma=0)}$ parameter in
Model $\textmd{II}$. Here, we can appreciate slight deviations
from the standard model (or equally the
$\Delta_{(\gamma=0)}\rightarrow0$ limit). This possibility does
not rule out the standard model, and may in principle be used to
distinguish between the $\Lambda CDM$ and our models.

\begin{figure}[!htbp]
\includegraphics[width=8cm,height=8cm]{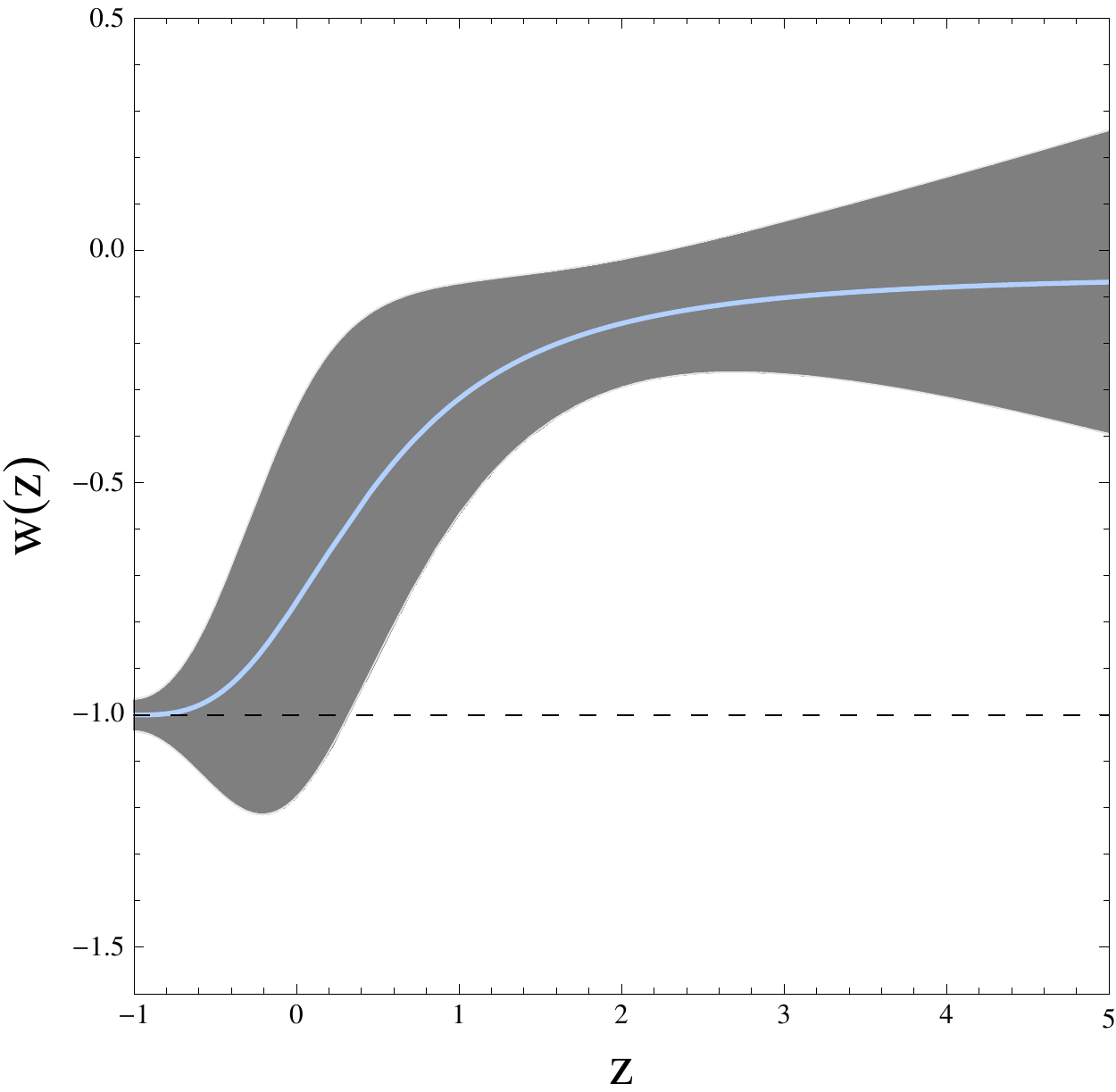}
\includegraphics[width=8cm,height=8cm]{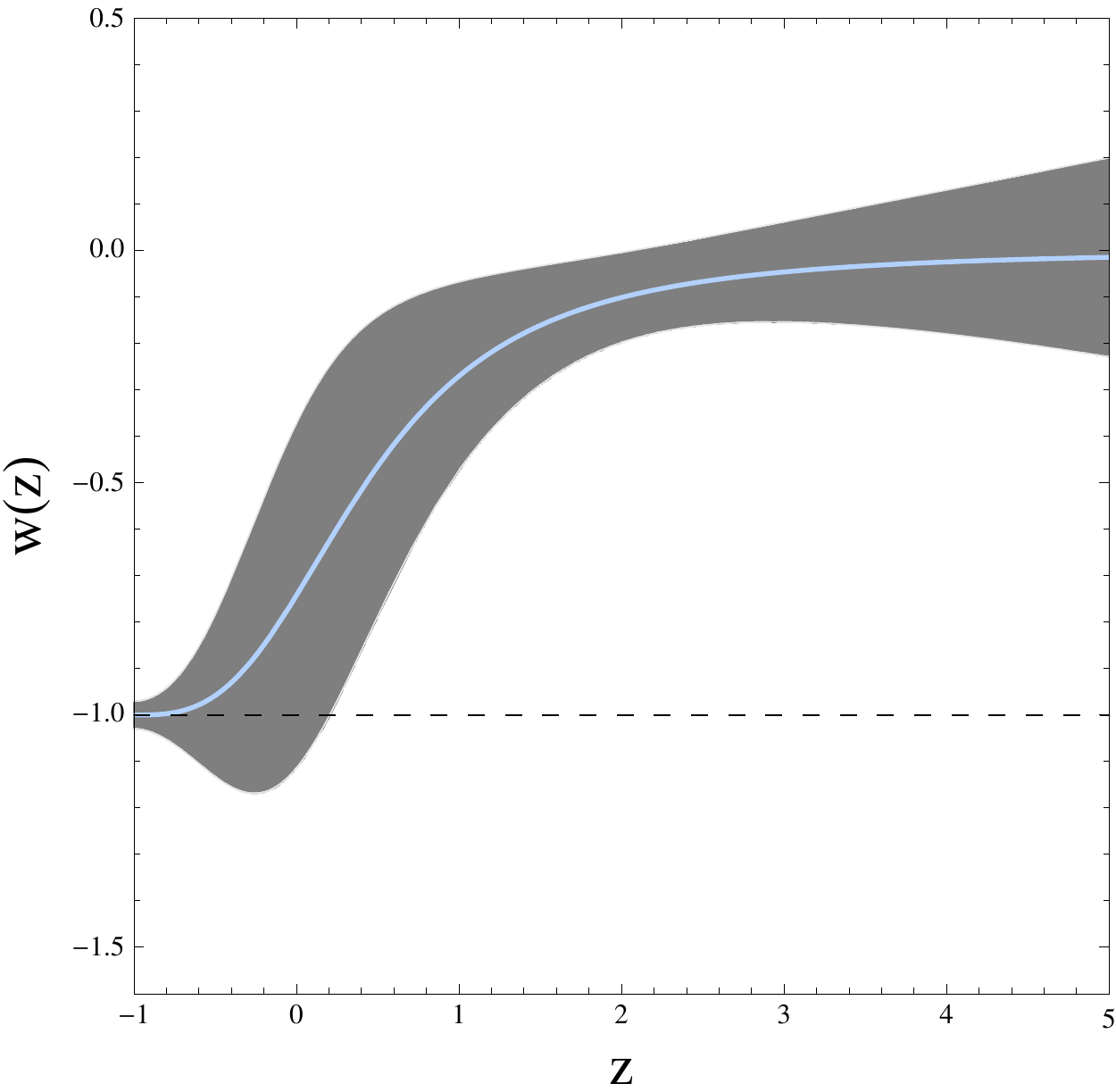}
\caption{Reconstruction of the EoS at 68\% CL (gray region) from
our joint analysis Model I (top panel) and Model II ($\gamma =0$)
(bottom panel). The blue line represents the best fit value for
all data set $SNIa + BAO + H(z)$.}
    \label{Weff}
\end{figure}

The equation of state (EoS) considering a given cosmological model
can be written as

\begin{equation}\label{42}
w(z) =-1-\frac{2}{3} \frac{\dot{H}}{H^2} = -1+ \frac{2(1+z)}{3H} \frac{dH}{dz},
\end{equation}

\noindent which has been derived from the combination of
Eqs.~(\ref{fff10}) and~(\ref{fff1}), for the general case and
Eq.~(\ref{mfe}) for the particular case of $\gamma = 0$.
Fig.~\ref{Weff}. shows the behavior of the total EoS for both
cases, with error propagation at 68.27\% CL regarding the best fit
values presented in table \ref{Om_del_gam}. We note that the total
EoS, due to the mechanism presented in this paper, does not cross
the phantom division line for the best fit of parameters. To high
redshift approach asymptotically to a value of zero, that is,
behaving like a fluid without pressure. In general, we note the
behavior $-1 \lesssim w \lesssim 0$ from the best fit values.
Similar behavior are found in unification models in the dark
sector of the universe.

On the other hand, it is natural to describe the kinematics of the
cosmic expansion through the Hubble parameter $H(t)$, and its
dependence on time, i.e. the deceleration parameter $q(z)$. The
deceleration parameter is defined as $q(z)=-\ddot{a}a/\dot{a}$
combined with the $\ddot{a}/a=H^2+\dot{H}$ relation, where
$\dot{H}=dH/dt$, to get

\begin{equation}\label{43}
q(z) = -1+\frac{(1+z)}{H(z)}\frac{dH(z)}{dz}.
\end{equation}

\noindent From Eq.~(\ref{refer4}) and by considering $p=0$ along
with $p^{\gamma}_e=-\rho^{\gamma}_e$, it follows that

\begin{equation}\label{44}
q_0 = \frac{\Omega_{m0}}{2} - \Omega^{\gamma}_e
\end{equation}

\noindent where its RHS has to be evaluated at $z=0$, and we have
defined $\Omega^{\gamma}_e=\rho^{\gamma}_e/\rho_{cr}$. In general,
if $\Omega^{\gamma}_e$ is sufficiently large (i.e.
$\Omega^{\gamma}_e >\Omega_m $), then $q(z=0)<0$, which
corresponds to an accelerated expanding universe.

\begin{figure}[!htbp]
\includegraphics[width=8cm,height=6cm]{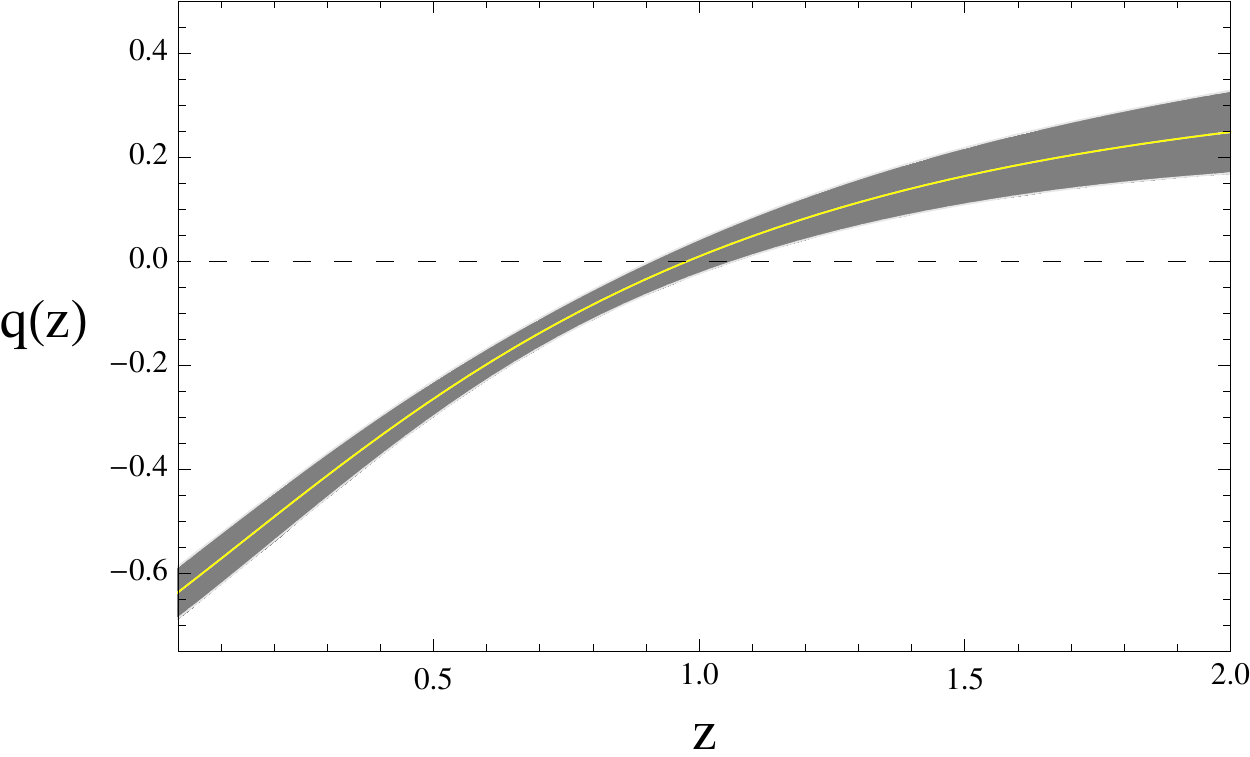}
\includegraphics[width=8cm,height=6cm]{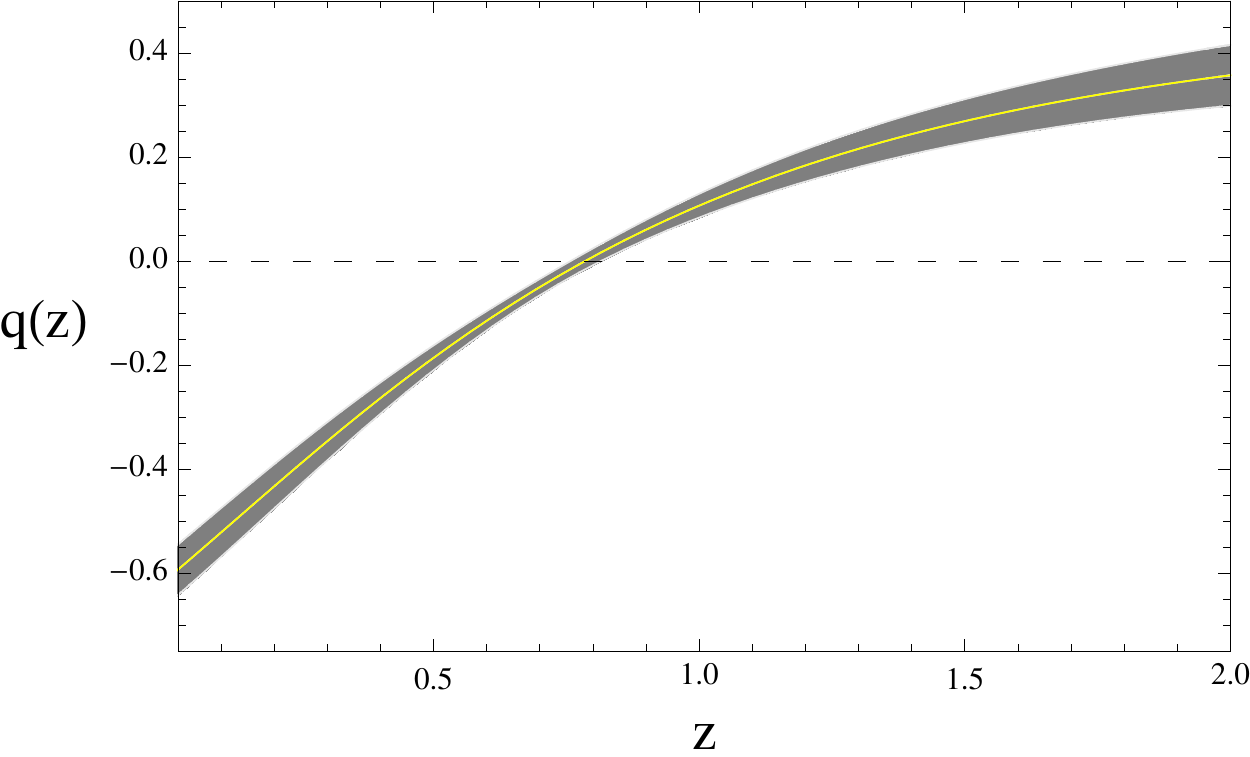}
\caption{Reconstruction of the $q(z)$ parameter, along with the
$1\sigma$ errors (shaded region). From our joint analysis Model I
(top panel) and Model II ($\gamma =0$) (bottom panel). The yellow
line represents the best fit value for all data set $SNIa + BAO +
H(z)$.} \label{qz}
\end{figure}

$q(z)$ has been plotted in Fig.~\ref{qz}. by using the best fit of
parameters with all observational data $SNIa+BAO+Hz$ (See Table
\ref{Om_del_gam}). As expected, models give $q(z)<0$ at late times
and $q(z)>0$ at earlier epoch, which means that the expansion rate
is slowed down in past and speeded up at present. Therefore, there
is a transition between decelerated phase ($q(z)>0$) into an
accelerated era $q(z)<0$ at redshift $z_t$ for these models. Our
analysis admits $\left\lbrace z_t = 0.98,q_0=-0.63\right\rbrace$
for Model $\textmd{I}$, and $\left\lbrace z_t =
0.77,q_0=-0.59\right\rbrace$ for Model $\textmd{II}$.

\begin{table*}[htb!]
\begin{center}
\begin{tabular}{cccccc}
\hline
\hline
Data                   &  $\chi^2_{min}$ & $\Delta$ & $\Omega_{m0}$          & $\gamma$ & $\Omega_C$ \\
\hline
SNIa                       & 562.227  & $0.001 \pm 1.702$   &  $0.279 \pm 0.033$     &  $0.001 \pm 0.031$ & $0.77\pm0.36$\\

SNIa+BAO             & 564.724  & $0.034 \pm 0.064$    & $0.289 \pm 0.015$      & $-0.021 \pm 0.018$ & $0.794\pm0.073$\\

SNIa+BAO+H(z)    & 583.613  & $0.183 \pm 0.053$    &  $0.274 \pm 0.013$     & $-0.012 \pm 0.016$  & $0.725\pm0.065$\\
\hline
Model $\textmd{II}$ & Data                       &  $\chi^2_{min}$ & $\Delta_{(\gamma=0)}$ & $\Omega_{m0}$          & $\Omega_{C_{(\gamma=0)}}$ \\
\hline
$\gamma=0$ &SNIa                       & 562.228  & $0.02 \pm 1.26$       &  $0.28 \pm 0.14$          &$0.72\pm0.15$ \\

$\gamma=0$ &SNIa+BAO             & 564.818  & $0.040 \pm 0.060$    & $0.287 \pm 0.013$      &$0.713\pm0.014$ \\

$\gamma=0$ &SNIa+BAO+H(z)    & 585.158  & $0.038 \pm 0.051$    &  $0.270 \pm 0.010$     &$0.729\pm0.010$ \\

\hline
\end{tabular}
\end{center}
\caption{Summary of the best fit values at 68.27\% CL for the
parameters
$\Delta(\equiv\frac{(6\gamma-1)\delta\pi}{(4\gamma-1)G})$,
$\Omega_{m0}$ and $\gamma$ to R\'{e}nyi entropy. Also, we shows
the summary of the best fit values at 68.27\% CL for the
particular case where $\gamma=0$, where
$\Delta=\Delta_{(\gamma=0)}$. We also present the value of
$\Omega_c$, derived from standard error propagation.}
      \label{Om_del_gam}
\end{table*}

\section{Summary and Concluding Remarks}

Applying the Clausius relation, the Cai-Kim temperature and the
R\'{e}nyi entropy to the apparent horizon of flat FRW universe, we
could get a model for the dynamics of universe. Fitting model to
observational data, the values of model parameters were obtained.
We found out that if we attribute R\'{e}nyi entropy to
horizon, then the current accelerated phase of the universe
expansion may be described in the Rastall framework. Our study
also shows that the probable non-extensive features of spacetime
may play the role of a varying dark energy in a universe in which
ordinary energy-momentum conservation law is valid.

Although our model shows a suitable agreement with observational
data, it is very important to study the evolution of density
perturbations in this model which helps us to decide about the
performance of our model. We leave this subject for the future
work.
\section*{Acknowledgment}
We thank the anonymous referee for his/her valuable comments and
suggestions. The authors thank to Rafael C. Nunes for helpful
discussions and critical reading of the manuscript. The work of H.
Moradpour has been supported financially by Research Institute for
Astronomy \& Astrophysics of Maragha (RIAAM) under research
project No. 1/5237-4. E. Abreu thanks CNPq (Conselho Nacional de
Desenvolvimento Cient\' ifico e Tecnol\'ogico), Brazilian
scientific support federal agency, for partial financial support,
Grants numbers 302155/2015-5 and 442369/2014-0 and the hospitality
of Theoretical Physics Department at Federal University of Rio de
Janeiro (UFRJ), where part of this work was carried out.

\end{document}